# Landau Level Splitting in Graphene in High Magnetic Fields

Y. Zhang[1], Z. Jiang[1,3], J. P. Small[1], M. S. Purewal[1], Y.-W. Tan[1], M. Fazlollahi[1], J. D. Chudow[1], J. A. Jaszczak[4], H. L. Stormer[1,2] and P. Kim[1]

[1]*Department of Physics and Department of Applied Physics, Columbia University, New York, New York 10027.*

[2]*Bell Labs, Lucent Technologies, Murray Hill, NJ 07974*

[3]*National High Magnetic Field Laboratory, Tallahassee, Florida 32310.*

[4]*Department of Physics, Michigan Technological University, Houghton, Michigan 49931.*

The quantum Hall (QH) effect in two-dimensional (2D) electrons and holes in high quality graphene samples is studied in strong magnetic fields up to 45 T. QH plateaus at filling factors $\nu = 0, \pm 1, \pm 4$ are discovered at magnetic fields $B > 20$ T, indicating the lifting of the four-fold degeneracy of the previously observed QH states at $\nu = \pm 4(|n|+1/2)$, where $n$ is the Landau level index. In particular, the presence of the $\nu = 0, \pm 1$ QH plateaus indicates that the Landau level at the charge neutral Dirac point splits into four sublevels, lifting sublattice and spin degeneracy. The QH effect at $\nu = \pm 4$ is investigated in tilted magnetic field and can be attributed to lifting of the spin-degeneracy of the $n = 1$ Landau level.

Graphene, a single atomic layer of graphite, is a monolayer of carbon atoms arranged in a hexagonal lattice. The conduction and the valence band in graphene touch at two inequivalent points (**K** and **K′**) at the corners of the Brillouin zone. Around those two points (termed "Dirac points"), the energy dispersion relation is linear and the electron dynamics appears thus "relativistic" where the speed of light is replaced by the Fermi velocity of graphene ($v_F \approx 10^6$ m/sec) [1-4]. Such a unique electronic band structure has profound implications for the quantum transport in graphene. Indeed, it has recently been observed that high mobility graphene samples exhibit an unusual sequence of quantum Hall (QH) effects [5, 6]. In a magnetic field, $B$, perpendicular to the graphene plane, the Landau levels (LL) have an energy spectrum $E_n = \text{sgn}(n)\sqrt{2e\hbar v_F^2 |n| B}$, where $e$ and $\hbar$ are electron charge and Plank's constant, and the integer $n$ represents an electron-like ($n>0$) or a hole-like ($n<0$) LL index. The appearance of an $n=0$ LL at the Dirac point indicates a special electron-hole degenerate LL due to the exceptional topology of the graphene band structure. Of particular interest are the QH states near the Dirac point where strong electron correlation may affect the stability of this single-particle LL due to many-body interaction [7].

In this letter, we present results from magnetotransport measurements in graphene samples in magnetic field up to 45 T. New QH states corresponding to a lifting of the spin- and sublattice symmetry- degenerate Landau levels are observed in $B>20$ T. These QH states are particularly well resolved in the vicinity of the Dirac point.

Single atomic layers of graphite are extracted from bulk graphite crystals [8] and deposited onto $SiO_2$/Si substrate using the mechanical method described in previous work [9]. Typically, graphene pieces of lateral size ~ 3-10 $\mu$m are chosen for device fabrication. Multiple electrodes arranged in Hall-bar geometry, are fabricated on the sample using electron beam lithography followed by Au/Cr (30/3 nm) evaporation and a lift-off process (Fig.1 left inset). The degenerately doped silicon substrate serves as a gate electrode with 300 nm thermally grown silicon oxide acting as the gate dielectric. By applying a gate bias voltage, $V_g$, the charge density of the sample can be tuned. The three graphene devices used in this experiment have mobilities as high as ~ $5 \times 10^4$ cm$^2$/Vs. The magneto-resistance ($R_{xx}$) and Hall resistance ($R_{xy}$) of the graphene sample are measured using Lock-in amplifiers at an excitation current of 10 nA in helium vapor in a cryostat.

Fig. 1 displays $R_{xx}$ and $R_{xy}$ measured simultaneously as a function of $V_g$ at $B=45$ T and at a temperature of $T=1.4$ K. A series of fully developed QH states, i.e., zeros in $R_{xx}$ and plateaus in $R_{xy}$ quantized to values $h/(e^2 \nu)$ with an integer filling factor $\nu$, are observed. Well-defined $\nu = \pm 2$ QH states and a $\nu = -6$ state are visible in these traces, in accordance with the previous low-magnetic field measurement ($B<9$ T) [5, 6]. In addition to these QH states, new QH states at $\nu = \pm 1$ and at $\nu = \pm 4$ are clearly resolved in the high magnetic field data, which is the central finding of this letter. We also note that both $R_{xx}$ and $R_{xy}$ fluctuate but remain at finite values (of the order of 100 k$\Omega$, see Fig.1 right inset) near the Dirac point where the carrier density is low [10].

In order to investigate the development of the new QH states in high magnetic fields, we carried out magnetotransport measurements in various magnetic fields between 9 T and 45 T at $T = 1.4$ K. The Hall conductivity, $\sigma_{xy}$, deduced from $R_{xx}$ and $R_{xy}$ [11], is plotted as a function of $V_g$ in Fig. 2. In addition to the QH plateaus at $\nu = \pm 4(|n| + \frac{1}{2})$ observed in lower magnetic field ($B < 9$ T), new QH plateaus appear at higher magnetic fields. Specifically, $\nu = 0$ QH plateau is resolved at $B > 11$ T, and $\nu = \pm 1, \pm 4$ plateaus appear at similar field $B > 17$ T. The $\nu = 0$ QH plateau right at the Dirac point ($V_g \approx 3.7$ V) is intriguing. While $\sigma_{xy}$ exhibits a clearly resolved $\nu = 0$ plateau at for B > 11 T (Fig. 2), $R_{xx}$ shows a finite peak ($R_{xx} \sim 40\,\text{k}\Omega$ as $|R_{xy}| \to 0$ at $B = 25$ T, see Fig. 2, upper inset). This new type of QH state does not conform to the standard QH observation (i.e., zeros in $R_{xx}$), and certainly deserves further study.

The new set of QH plateaus, $\nu = 0, \pm 1$ and $\pm 4$, can be attributed to the magnetic field induced splitting of the $n = 0$ and $n = \pm 1$ LLs. In lower magnetic fields, each LL at energy $E_n$ is assumed to be four-fold degenerate due to a two-fold spin degeneracy and a two-fold sublattice symmetry (i.e. **K** and **K**′ degeneracy) [12]. The observed filling factor sequence $\nu = 0$ ($B >$ 11 T) and $\nu = \pm 1$ ($B >$ 17 T) implies that the degeneracy of $n = 0$ are fully lifted at high magnetic fields, such that $\sigma_{xy}$ increases in steps of $e^2/h$ as the Fermi energy, $E_F$, passes through the LLs, whose substructure is now resolved. While the $n = 0$ LL is totally resolved into $\nu = 0, \pm 1$ plateaus the four-fold degeneracy in the $n = \pm 1$ LLs is only partially resolved into $\nu = \pm 4$ leaving a two-fold degeneracy in each of these LLs [13]. Although a close examination of $R_{xx}$ and $R_{xy}$ hints at developing QH states corresponding to $\nu = \pm 3$ (Fig. 1, gray arrows), their relatively weaker features strongly indicate that there exists a hierarchy of degeneracy lifting in these LLs as shown schematically in the lower left inset of Fig. 2.

Considering that the presence of a magnetic field alone does not break the inversion symmetry of graphene lattice, the broken sublattice symmetry in graphene in high magnetic fields is rather surprising. The observed LL splitting bears a resemblance to the valley degeneracy splitting observed in the 2D electron gas in Si inversion layers [14] and AlAs quantum wells [15], where the lifting of the valley degeneracy occurs due to uniaxial strain or due to a many-body induced correlation effect such as "valley skyrmions" [16]. Since the sublattice symmetry is protected by inversion symmetry in graphene, simple uniaxial strain does not lift the graphene sublattice degeneracy, leaving many-body electron correlation within the LL as an alternative origin for the lifting of the degeneracy at the Dirac point. We point out that a high magnetic field anomaly has been observed in bulk graphite [17], possibly originating from the formation of a charge density wave (CDW) due to Fermi surface nesting of the electron and hole LLs. In addition, a recent theoretical study suggests that CDW order may open up a gap at the Dirac point in high magnetic fields [18], further substantiating the possibility of a many-body origin of our observation of fully lifted LL degeneracies.

We now turn our attention to $\nu = \pm 4$ QH states and study the origin of the lifting of the degeneracy in the $n = \pm 1$ LLs. We first determine the energy gap between LLs, $\Delta E$, by measuring the temperature dependence of the associated $R_{xx}$ minimum, $R_{xx}^{\min}$. The inset of Fig. 3 shows Arrhenius plots of $R_{xx}^{\min}$ which reveal thermally activated behavior, $R_{xx}^{\min} \sim \exp[-\Delta E / 2k_B T]$ ($k_B$ is the Boltzmann constant). We find that $\Delta E$, derived from linear fits to these data, depends linearly on the magnitude of total magnetic field $B_{tot}$ (the main section of Fig. 3). The energy gap at $\nu = \pm 4$ could be the result of two possible scenarios: (i) Spin splitting whose energy scales as $\Delta E_S = g^* \mu_B B_{tot}$, where $g^*$ is the effective Lande $g$-factor and $\mu_B$ the Bohr magneton; or alternatively, (ii) sublattice symmetry breaking and gap formation due to many-body correlations. In order to distinguish these two different mechanisms, we carried out transport measurement in tilted magnetic field. Since the electron system in graphene is only ~ 3 Å thick, a total magnetic field, $B_{tot}$, applied at tilt angle, $\theta$, with respect to the direction normal to the graphene plane (Fig. 4a inset) affects in-plane motion relevant to electron-electron correlation only via the perpendicular field $B_p = B_{tot} \cos(\theta)$. The electron spin, on the other hand, experiences the full $B_{tot}$, and hence spin splitting is independent of $\theta$. Fig. 4(a) shows the minima in $R_{xx}$ corresponding to the $\nu = -4$ QH state at identical $B_p$ but at different $B_{tot}$: $B_{tot} = 30$ T ($\theta = 0°$) and $B_{tot} = 45$ T ($\theta = 49.0°$). The fact that at fixed $B_p$, $R_{xx}^{\min}$ depends on $B_{tot}$ strongly suggests that $\nu = -4$ originates from spin splitting of the $n = -1$ LL and not from any in-plane electron correlation. Following this argument, the energy gap of these spin-split LLs can now be written as [14]

$$\Delta E = \Delta E_S - 2\Gamma \qquad (1)$$

where $\Gamma$ is the half-width of the LL broadening at half maximum. Assuming that $\Gamma$ is constant, then $g^*$ and $\Gamma$ can be extracted from the slope and y-intercept of linear fit: $g^* = 2.0$ and $\Gamma = 18.2\,\text{K}$ are obtained from Fig. 3 for $\nu = -4$ QH state and $g^* = 1.7$ and $\Gamma = 15.8\,\text{K}$ for $\nu = 4$ QH state from a similar analysis (not shown). The fact that these measured g-factors are very close to that of the bare electron ($g^* = 2$) further substantiates the spin-splitting origin of the $\nu = \pm 4$ QH states. We also note that these $\Gamma$ values are in good agreement with the estimation, $\Gamma \sim \hbar/\tau$, where $\tau \approx 100$ fs is the relaxation time deduced from the carrier mobility in the zero magnetic field limit.

Finally, we discuss our measurement of $R_{xx}$ at $\nu = \pm 4$ at $T = 1.4$ K in a tilted magnetic field with a total fixed magnitude. Here we kept $B_{tot} = 45$ T, and varied $\theta$ to change $B_p$. As shown in Fig. 4(b), the $R_{xx}$ minima display pronounced changes as $B_p$ increases, seemingly contradicting our previous conclusion that $\nu = -4$ QH states originates from spin polarized LLs whose energy level separation ($\Delta E_S$) only depends on $B_{tot}$. This contradiction is resolved if we relax our previous assumption regarding a constant $\Gamma$, and take into account the $B_p$ dependence of the Landau level width $\Gamma(B_p)$. Indeed, a phenomenological linear relation for $\Gamma(B_p)$ can be obtained from the exponential dependence of $R_{xx}^{\min}$ on $B_p$ (inset to Fig. 4(b)). From the linear fit (solid line in the inset to Fig. 4(b)), we obtain $\Gamma(B_p) = \Gamma_0(\text{K}) - 0.14 B_p(\text{T})$ up to a constant, $\Gamma_0$, the LL broadening at the zero field limit. Combining this functional form of $\Gamma(B_p)$ with Eq. (1), the experimentally observed linear relation of $\Delta E(B_{tot} = B_p)$ (Fig. 3) now yields slightly reduced g-factors: $g^* = 1.8$ for $\nu = -4$ and $g^* = 1.4$ for $\nu = +4$.

In summary, we report the observation of splittings of LLs under high magnetic field up to 45 T. We discover the $n = 0$ LL splits into four sublevels, lifting spin and sublattice degeneracy, potentially indicating a many-body correlation in this LL. In the $n = \pm 1$ LL, only the spin degeneracy seems to be lifted within experimentally accessible high magnetic field. The effective $g$-factors were found close to the bare electron $g$-factor.

We thank A. Millis, C. Kane, E. Mele, M. Shayegan, M. P. A. Fisher, F. D. Haldane, A. H. C. Neto, and J. Zhu for helpful discussions. This work is supported by the DOE (DE-FG02-05ER46215 and DE-AIO2-04ER46133) and NSF under DMR-03-52738. A portion of this work was performed at the National High Magnetic Field Laboratory, which is supported by NSF Cooperative Agreement No. DMR-0084173, by the State of Florida, and by the DOE.

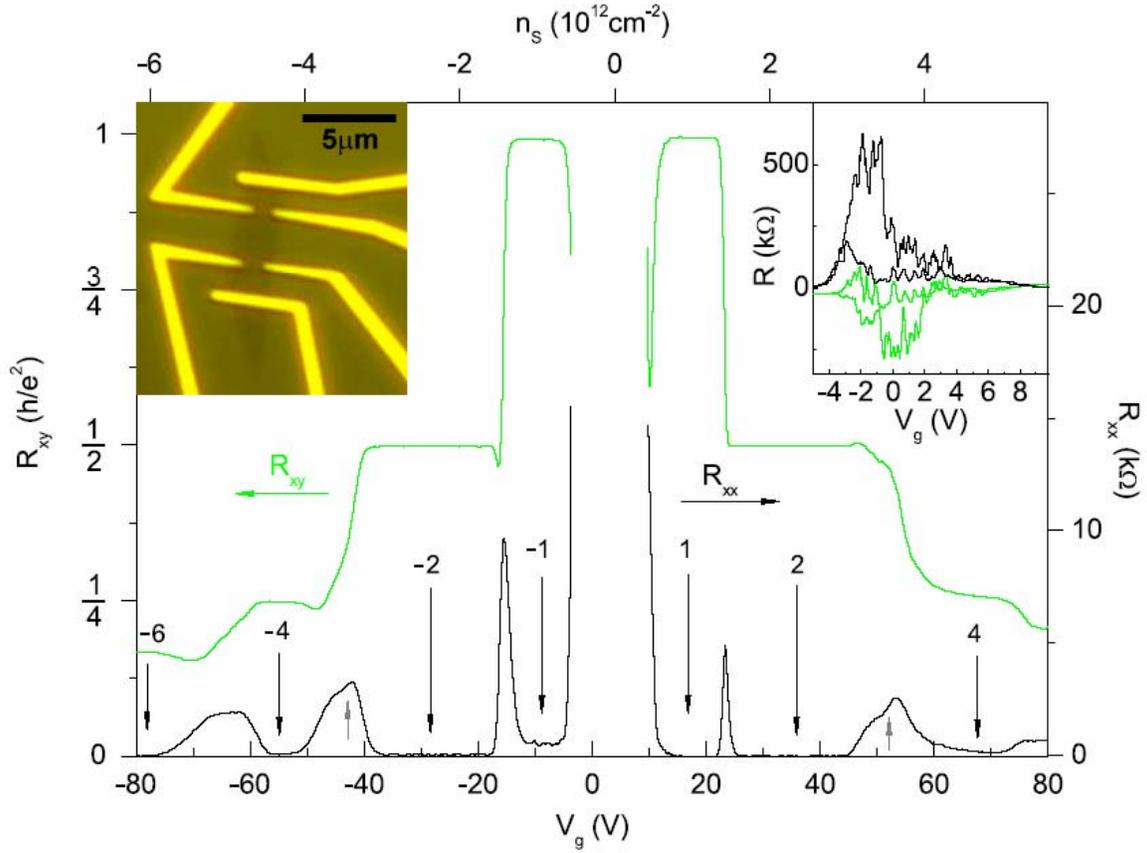

**FIG 1.** (color online) $R_{xx}$ and $R_{xy}$ measured in the device shown in the left inset, as a function of $V_g$ at $B = 45$ T and $T = 1.4$ K. $-R_{xy}$ is plotted for $V_g > 0$. The numbers with the vertical arrows indicate the corresponding filling factor $\nu$. Gray arrows indicate developing QH states at $\nu = \pm 3$. $n_s$ is the sheet carrier density derived from the geometrical consideration. Right inset: $R_{xx}$ (dark solid lines) and $R_{xy}$ (light solid lines) for another device measured at $B = 30$ T and $T = 1.4$ K in the region close to the Dirac point. Two sets of $R_{xx}$ and $R_{xy}$ are taken at different time under the same condition. Left inset: an optical microscope image of a graphene device used in this experiment.

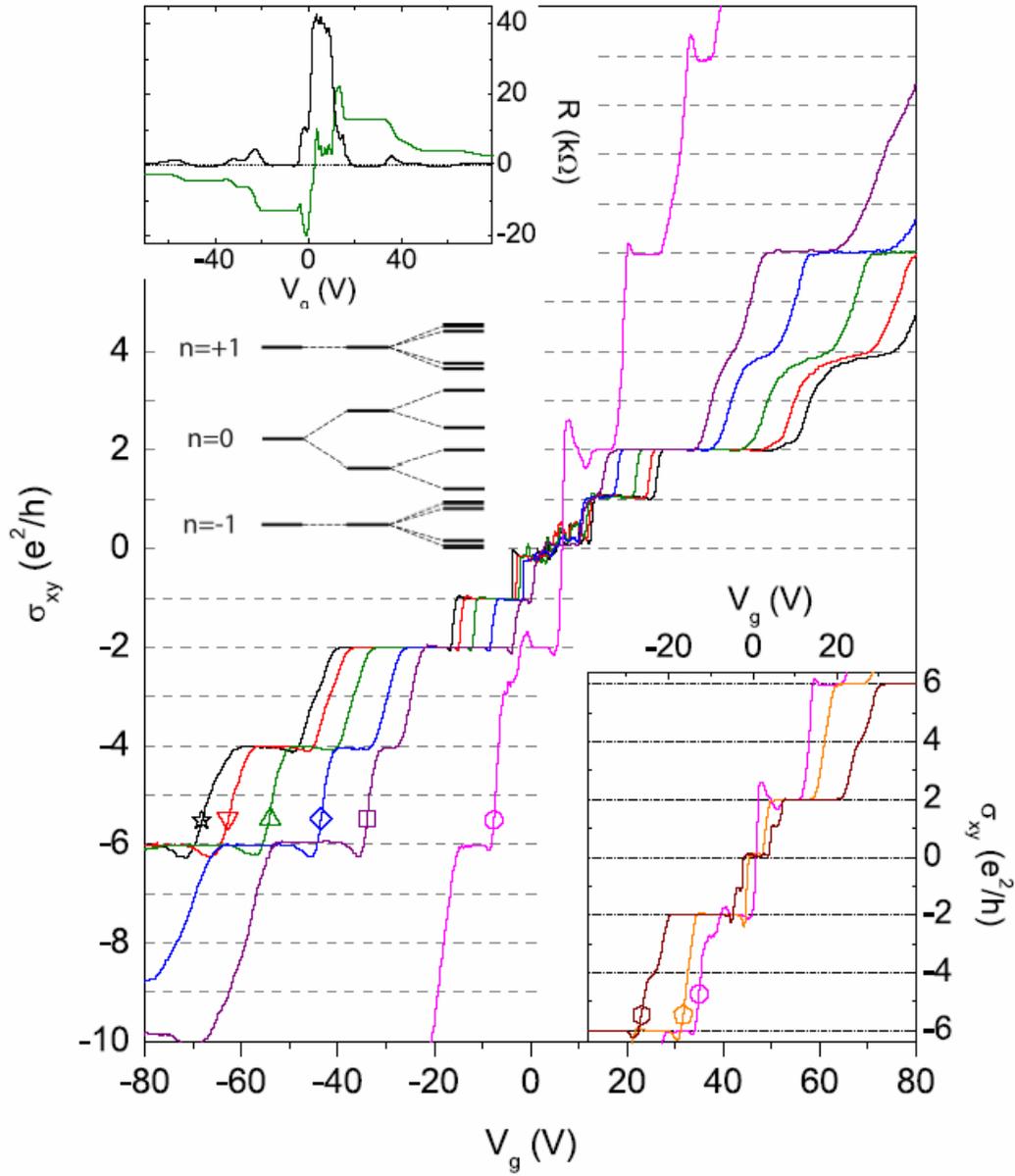

**FIG 2.** (color online) $\sigma_{xy}$, as a function of $V_g$ at different magnetic fields: 9 T (circle), 25 T (square), 30 T (diamond), 37 T (up triangle), 42 T (down triangle), and 45 T (star). All the data sets are taken at $T = 1.4$ K, except for the $B = 9$ T curve, which is taken at $T = 30$ mK. Left upper inset: $R_{xx}$ and $R_{xy}$ for the same device measured at $B = 25$ T. Left lower inset: a schematic drawing of the LLs in low (left) and high (right) magnetic field. Right inset: detailed $\sigma_{xy}$ data near the Dirac point for B = 9 T (circle), 11.5 T (pentagon) and 17.5 T (hexagon) at $T = 30$ mK.

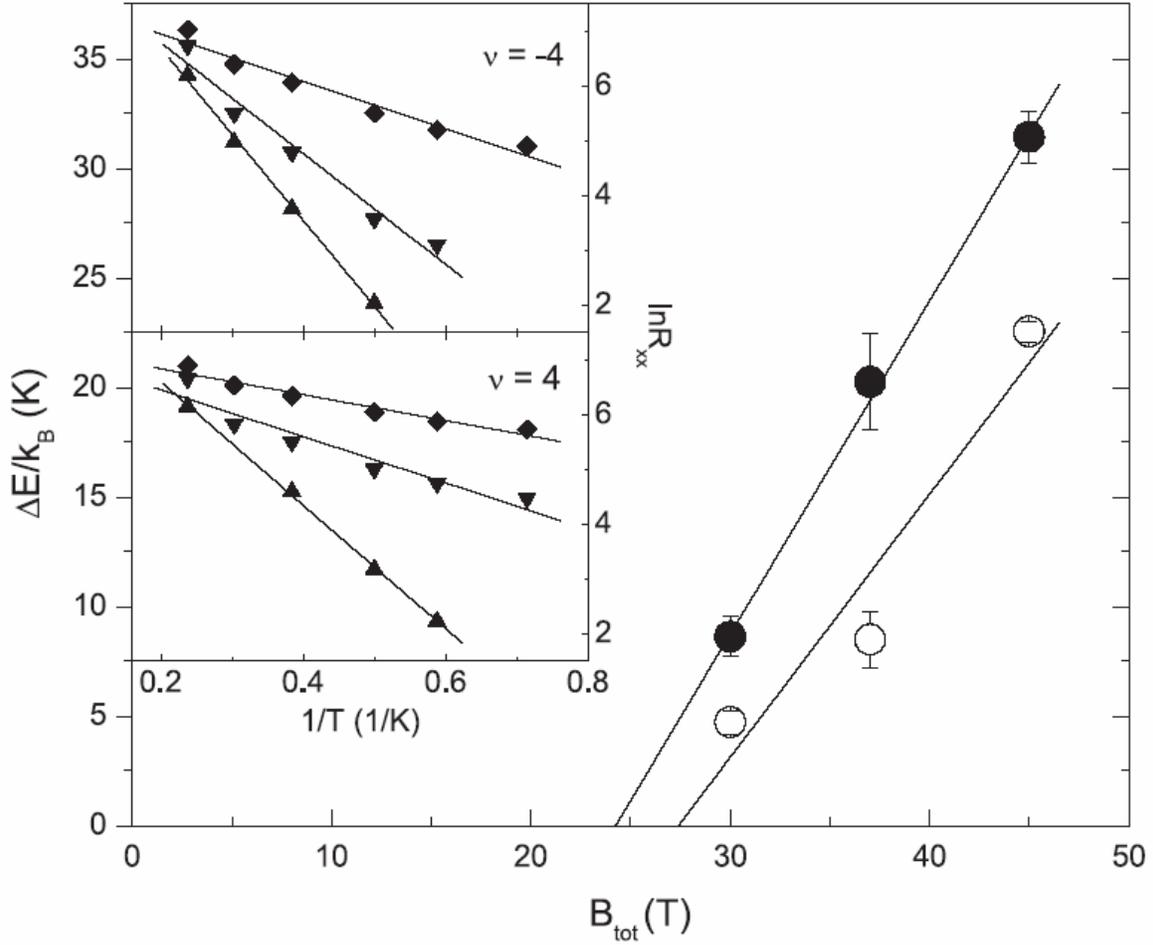

**FIG 3.** Inset: $R_{xx}^{min}$ at $\nu = -4$ (upper inset) and $\nu = 4$ (lower inset) for $B_{tot} = 30$ T (diamond), 37 T (down triangle) and 45 T (up triangle), applied to normal to graphene plane. The straight lines are linear fits to the data. The main figure: the activation energy, $\Delta E$, as a function of $B_{tot}$ for filling factor $\nu = -4$ (filled circle) and $\nu = +4$ (open circle). The straight lines are linear fits to the data.

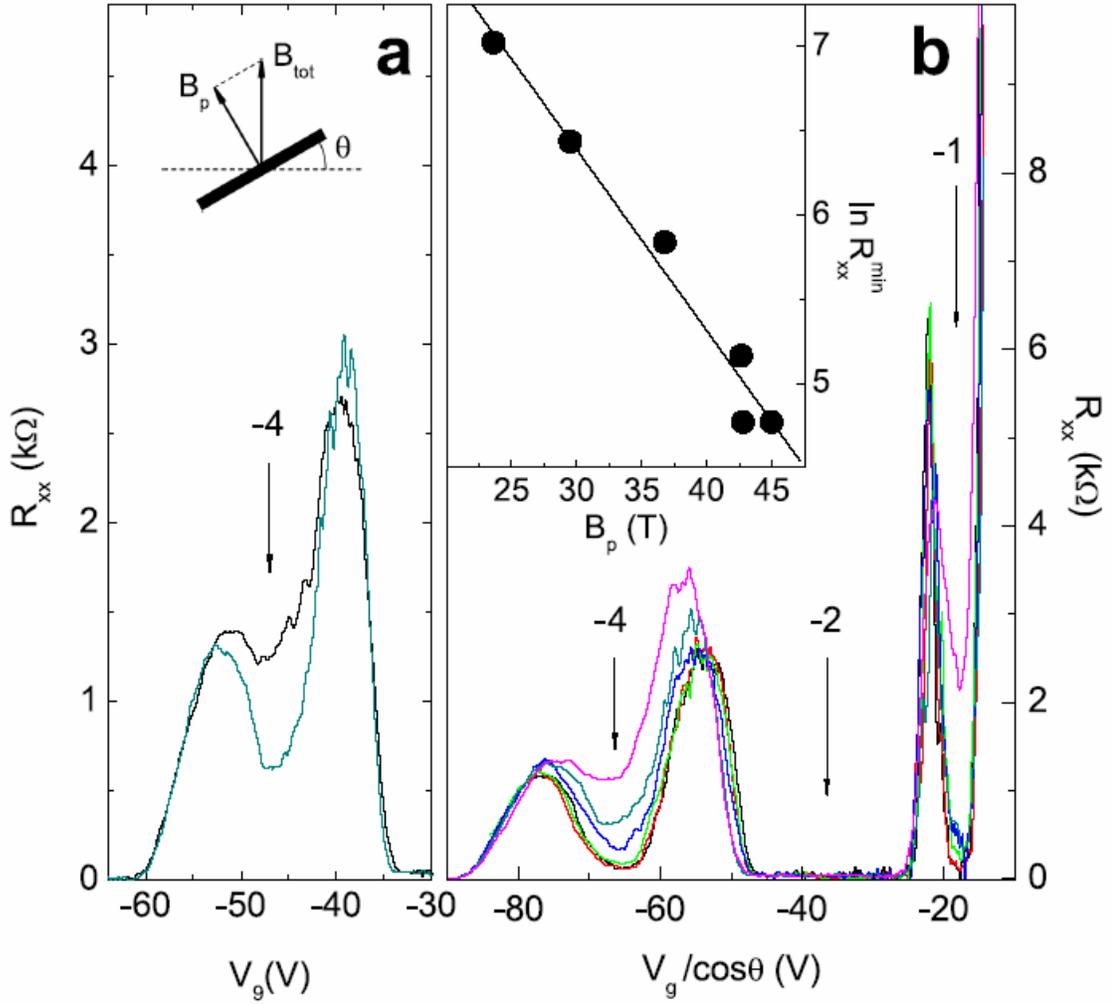

**FIG 4.** (color) (a) $R_{xx}$ measured as a function of gate voltage around $\nu = -4$ at $B_p = 30$ T at two different total magnetic fields, $B_{tot} = 45$ T (light solid line) and $B_{tot} = 30$ T (dark solid line). Numbers with vertical arrow indicates filling factors. Inset: a schematic drawing of the measurement configuration. (b) $R_{xx}$ measured as a function of $V_g$. Data are taken at $B_{tot} = 45$ T at six different tilt angles: $\theta = 0°$, 18.1°, 18.8°, 35.5°, 49.0°, 58.4° (ascending order from bottom at $\nu = -4$). Inset: $R_{xx}^{min}$ at $\nu = -4$ plotted on natural logarithmic scale as a function of $B_p$. The straight line is a linear fit to the data.